\documentstyle[twocolumn,aps]{revtex}

\begin{document}
\title{Non-uniform thermal magnetization noise in thin films: application to GMR
heads}
\author{Vladimir L. Safonov and H. Neal Bertram}
\address{Center for Magnetic Recording Research, University of California - San\\
Diego, 9500 Gilman Drive, La Jolla, CA 92093-0401}
\date{\today}
\maketitle

\begin{abstract}
A general scheme is developed to analyze the effect of non-uniform thermal
magnetization fluctuations in a thin film. The normal mode formalism is
utilized to calculate random magnetization fluctuations. The magnetization
noise is proportional to the temperature and inversely proportional to the
film volume. The total noise power is the sum of normal mode spectral noises
and mainly determined by spin-wave standing modes with an odd number of
oscillations. The effect rapidly decreases with increasing mode number. An
exact analytical calcutaion is presented for a two-cell model.
\end{abstract}


\section{Introduction}

Recently thermal magnetization fluctuations have been shown to be an
important issue for thin film magnetoresistive heads used for high-density
magnetic recording [1]. Noise estimates have been initially made for the
simplest model of a uniformly magnetized film [1-3]. In order to develop a
more realistic model it is necessary to analyze the role of non-uniform
magnetization fluctuations. In this paper we propose a general scheme to
calculate the effect of non-uniform thermal magnetization fluctuations in an
ultrathin film. After discretizing the film into magnetic cells, the
well-known normal mode formalism is utilized. By doing so we represent the
magnetization noise in terms of independent standing spin waves subjected to
white thermal noise. As an example, the explicit solution is given for a
two-mode model.

\section{Calculation scheme}

Let us consider a thin film of the volume $V=d_{x}\times d_{y}\times d_{z}$,
where $d_{x}\ll d_{y}$, $d_{z}$ are the thickness, width and length,
respectively. The thickness is assumed to be of the order or less than
exchange length ($d_{x}\leq \sqrt{A}/M_{{\rm s}}$) and therefore non-uniform
magnetization dynamics along the $x$ direction will be neglected. The film
is discretized in the $yz$ plane into $N$ identical magnetic cells. Each
cell is assumed to be uniformly magnetized and can be represented as a
classical spin ${\bf S}_{j}=-{\bf M}({\bf r}_{j})V_{{\rm c}}/\hbar \gamma $,
where ${\bf M}({\bf r}_{j})$ is the magnetization in the cell ($|{\bf M}(%
{\bf r}_{j})|=M_{{\rm s}}$), $V_{{\rm c}}=V/N$ is the cell volume, $M_{{\rm s%
}}$ is the saturation magnetization and $\gamma $ is the gyromagnetic ratio.
Planck's constant $\hbar $ is used for convenience as a dimensional
variable. Our approach is purely classical and final result does not contain 
$\hbar $.

The energy of the spin system can be written as 
\begin{equation}
{\cal E}={\cal E}_{{\rm anis}}+{\cal E}_{{\rm Z}}+{\cal E}_{{\rm dd}}+{\cal E%
}_{{\rm exch}},  \label{hamiltonian}
\end{equation}
where

\begin{equation}
{\cal E}_{{\rm anis}}=\frac{\hbar \gamma }{2S}\sum_{j=1}^{N}\left[ -H_{{\rm K%
},j}^{({\rm in})}({\bf n}_{j}\cdot {\bf S}_{j})^{2}+H_{{\rm K},j}^{(\perp
)}(S_{j}^{x})^{2}\right]  \label{h-anis}
\end{equation}
is the energy of anisotropy, ${\bf n}_{j}$ is the unit vector along the axis
of in-plane uniaxial anisotropy, $H_{{\rm K},j}^{({\rm in})}$ is the local
uniaxial\ anisotropy field, $H_{{\rm K},j}^{(\perp )}$ is the local
hard-axis anisotropy field.

\begin{equation}
{\cal E}_{{\rm Z}}=\hbar \gamma \sum_{j=1}^{N}{\bf H}_{j}\cdot {\bf S}_{j}
\label{h-Zeem}
\end{equation}
is the Zeeman energy with local magnetic fields ${\bf H}_{j}$. ${\cal E}_{%
{\rm dd}}$ is the energy of magnetostatic interactions between cells and $%
{\cal E}_{{\rm exch}}$ describes the exchange interaction between the
nearest neighboring cells. Boundary conditions can be taken into account by
choosing corresponding anisotropy fields.

The calculation procedure consists of the following steps. From (\ref
{hamiltonian}) it is necessary to find the equilibrium direction in the $yz$
plane for each spin (cell). This direction is defined by the angle $\theta
_{j}$ relative to $z$ axis. Then we represent the spin in its local
``quantization axes'' ${\bf x}_{j}$,{\bf \ }${\bf y}_{j}$, ${\bf z}_{j}$,
which are associated with the spin equilibrium direction (${\bf S}%
_{j}\parallel {\bf z}_{j}$):

\begin{eqnarray}
S_{j}^{y} &=&S_{j}^{y_{j}}\cos \theta _{j}+S_{j}^{z_{j}}\sin \theta _{j},
\label{coord} \\
S_{j}^{z} &=&S_{j}^{z_{j}}\cos \theta _{j}-S_{j}^{y_{j}}\sin \theta
_{j},\quad S_{j}^{x}=S_{j}^{x_{j}}.  \nonumber
\end{eqnarray}
Now we analyze small deviations for each spin from equilibrium. It is
convenient to describe small magnetization oscillations by conventional
spin-wave technique. Using a linearized Holstein-Primakoff transformation 
\cite{hopri}, we introduce complex variables$\;a^{\ast },\,a$, which are
classical analogs of creation and annihilation operators:

\begin{eqnarray}
S_{j}^{z_{j}} &=&-S+a_{j}^{\ast }a_{j},\quad S_{j}^{y_{j}}\simeq
(a_{j}+a_{j}^{\ast })\sqrt{2S}/2,  \label{h-pa} \\
S_{j}^{x_{j}} &\simeq &(a_{j}-a_{j}^{\ast })\sqrt{2S}/2i.  \nonumber
\end{eqnarray}
The quadratic part of the energy ({\ref{hamiltonian}}) can be written in the
form:

\begin{equation}
{\cal E}^{(2)}/\hbar =\sum_{k,l}\left( A_{kl}a_{k}^{\ast }a_{l}+\frac{B_{kl}%
}{2}a_{k}^{\ast }a_{l}^{\ast }+\frac{B_{kl}^{\ast }}{2}a_{k}a_{l}\right) ,
\label{quadform}
\end{equation}
where $\,A_{kl}=A_{lk}^{\ast }$ and $B_{kl}=B_{lk}$.

To eliminate the nondiagonal terms from (\ref{quadform}) we shall use the
linear canonical transformation \cite{tyablikov}: 
\begin{equation}
a_{j}=\sum_{k=1}^{N}\left( u_{jk}c_{k}+v_{jk}^{\ast }c_{k}^{\ast }\right) ,
\label{canonical}
\end{equation}
where $c_{k}$ and $c_{k}^{\ast }$ describe complex variables of normal
spin-wave modes. From gyromagnetic equations we have the algebraic equations
for eigenvalues $\omega _{k}$ (standing spin wave frequencies) and
eigenvectors $u_{lk}$, $v_{lk}$:

\begin{eqnarray}
\omega _{k}u_{jk} &=&\sum\limits_{l=1}^{N}{(}A_{jl}u_{lk}+B_{jl}v_{lk}), 
\nonumber \\
-\omega _{k}v_{jk} &=&\sum\limits_{l=1}^{N}{(}A_{jl}^{\ast
}v_{lk}+B_{jl}^{\ast }u_{lk}).  \label{system}
\end{eqnarray}
This system must be supplemented by the orthogonality and normalization
conditions:

\begin{eqnarray}
\sum\limits_{j=1}^{N}(u_{jl}u_{jk}^{\ast }-v_{jl}v_{jk}^{\ast }) &=&\delta
_{lk},~\sum\limits_{j=1}^{N}\left( u_{jl}v_{jk}{-}u_{jk}v_{jl}\right) =0, 
\nonumber \\
\sum\limits_{k=1}^{N}(u_{jk}u_{lk}^{\ast }-v_{jk}v_{lk}^{\ast }) &=&\delta
_{jl},~\sum\limits_{k=1}^{N}\left( u_{jk}v_{lk}{-}u_{jk}v_{lk}\right) =0.
\label{ortho}
\end{eqnarray}
The resulting quadratic energy form describes $N$ independent normal modes
(harmonic oscillators): 
\begin{equation}
{\cal E}^{(2)}/\hbar =\sum_{k=1}^{N}\omega _{k}c_{k}^{\ast }c_{k}.
\label{ener}
\end{equation}

The evolution of each normal spin-wave mode is characterized by frequency ($%
\omega _{k}$) and relaxation rate ($\eta _{k}$): $c_{k}\propto \exp
(-i\omega _{k}t-\eta _{k}t)$. From equipartition, white thermal noise ($k_{%
{\rm B}}T$) is applied to each independent mode. Using the solution for a
damped harmonic oscillator with additive noise (see, \cite{safbertnoise}, 
\cite{BJS}), we calculate the spectral density of the magnetic fluctuations:

\begin{eqnarray}
S_{c_{k}^{\ast }c_{l}}(\omega ) &=&\int_{-\infty }^{\infty }\langle
c_{k}^{\ast }(t)c_{l}(0)\rangle e^{-i\omega t}dt  \nonumber \\
&=&\delta _{kl}\frac{2(\eta _{k}/\omega _{k})}{(\omega _{k}-\omega
)^{2}+\eta _{k}^{2}}\frac{k_{{\rm B}}T}{\hbar },  \label{spectral}
\end{eqnarray}
where $T$ is the temperature, $\langle ...\rangle $ means thermal averaging.
Note that the total system has thermal energy $Nk_{{\rm B}}T$.

\section{Application to a MR head}

As a general application to a GMR head we take the $z$ direction to be the
cross track direction along which the current is applied and thus the
deviations of magnetization in the $y$ direction are responsible for
magnetoresistance. The power spectral density due to random magnetization
fluctuations can be written as \cite{smith},\cite{safbertnoise}:

\begin{equation}
S_{VV}(\omega )=C_{V}^{2}S_{m_{y}m_{y}}(\omega ),  \label{newpowerVV}
\end{equation}

\begin{equation}
C_{V}\equiv I_{{\rm bias}}(\partial R/\partial H_{{\rm ext}})(\partial H_{%
{\rm ext}}/\partial m_{y}),  \label{coef}
\end{equation}
where

\begin{equation}
S_{m_{x}m_{x}}(\omega )=\int_{-\infty }^{\infty }\langle \delta
m_{y}(t)\delta m_{y}(0)\rangle e^{i\omega t}dt.  \label{mag-noise}
\end{equation}
defines the magnetization noise. Here $\delta m_{y}$ denotes the net random
deviation of the normalized total magnetic moment: 
\begin{equation}
\delta m_{y}=-\sum_{j}^{N}\delta S_{j}^{y_{j}}\cos \theta _{j}/NS.
\label{dmy}
\end{equation}

Using Eqs. (\ref{h-pa}) and (\ref{canonical}), we rewrite $\delta m_{y}(t)$
in terms of the normal mode variables $c_{k}(t)$ and $c_{l}^{\ast }(t)$.
Thus, the correlation function (\ref{mag-noise}) can be rewritten in terms
of correlation functions $\langle c_{k}(t)c_{l}^{\ast }(0)\rangle $ and $%
\langle c_{k}^{\ast }(t)c_{l}(0)\rangle $. Utilizing (\ref{spectral}), we
finally obtain:

\begin{equation}
S_{m_{y}m_{y}}(\omega )=\frac{\gamma k_{{\rm B}}T}{M_{{\rm s}}V}%
\sum_{k=1}^{N}|W_{k}|^{2}\frac{\eta _{k}}{\omega _{k}}F_{k}(\omega ),
\label{final}
\end{equation}

\[
F_{k}(\omega )=\frac{1}{(\omega _{k}-\omega )^{2}+\eta _{k}^{2}}+\frac{1}{%
(\omega _{k}+\omega )^{2}+\eta _{k}^{2}}, 
\]

\begin{equation}
W_{k}=N^{-1/2}\sum_{j=1}^{N}(u_{jk}+v_{jk})\cos \theta _{j}.  \label{weight}
\end{equation}
Thus the magnetization noise power (\ref{final}) is proportional to the
temperature and inversely proportional to the film volume, and is the sum of
mode spectral functions $(\eta _{k}/\omega _{k})F_{k}(\omega )$ with some
``weights'' $W_{k}$ (\ref{weight}). In the case of $N=1$ we obtain the
result of \cite{safbertnoise},\cite{BJS}.

The summation of amplitudes of spin waves $u_{jk}+v_{jk}$ in (\ref{weight})
qualitatively means an integration of oscillations in the standing mode
(see, Fig.1). For example, if equilibrium is along the $z$ direction ($%
\theta _{j}=0$) the integration over the modes in Fig.1b and 1d vanish and
only modes from Fig.1a and 1c will contribute. Due to partial cancellation
in the odd spin-wave standing mode (see, e.g., Fig.1c) the weight rapidly
decreases with increasing mode wave number.

\section{Example: two mode model}

Let us consider now the case of two-cell discretization (Fig.2). We take
into account the ``easy'' anisotropy field $H_{{\rm K}}$ along the $z$ axis
and the ``hard'' axis field $4\pi M_{{\rm s}}$ perpendicular to the plane.
In this example only magnetostatic interaction between cells is included
taken in the form: $(\hbar \gamma )^{2}(\zeta {\bf S}_{1}\cdot {\bf S}%
_{2}-\zeta _{y}S_{1}^{y}S_{2}^{y}{\bf -}\zeta _{z}S_{1}^{z}S_{2}^{z})/V_{%
{\rm c}}^{3}$, where $\zeta $, $\zeta _{y}${\bf \ }and{\bf \ }$\zeta _{z}$
are the factors, which depend on the magnetostatic energy for a given film
geometry. For example, for a film of $4\times 180\times 250$ nm$^{3}$ $\zeta
\approx 0.23$, $\zeta _{y}\approx 0.17${\bf \ }and{\bf \ }$\zeta _{z}\approx
0.063$.

The energy quadratic form in terms of spin deviations (\ref{h-pa}) can be
written as:

\begin{eqnarray}
{\cal E}^{(2)}/\hbar &=&A(a_{1}^{\ast }a_{1}+a_{2}^{\ast
}a_{2})+B(a_{1}^{\ast }a_{2}+a_{2}^{\ast }a_{1})+C(a_{1}a_{2}  \nonumber \\
&+&a_{1}^{\ast }a_{2}^{\ast })+(D/2)(a_{1}a_{1}+a_{1}^{\ast }a_{1}^{\ast
}+a_{2}a_{2}+a_{2}^{\ast }a_{2}^{\ast }),  \label{twoE}
\end{eqnarray}
where

\begin{eqnarray*}
A &=&\gamma \{H_{{\rm K}}(\cos ^{2}\theta -%
{\frac12}%
\sin ^{2}\theta )-M_{{\rm s}}[\zeta -3(\zeta _{y}\sin ^{2}\theta \\
&&+\zeta _{z}\cos ^{2}\theta )]+2\pi M_{{\rm s}}+H_{0}\cos (\theta
_{H}-\theta )\}, \\
B &=&\gamma M_{{\rm s}}\zeta +C,\quad D=-\gamma (2\pi M_{{\rm s}}+H_{{\rm K}%
}\sin ^{2}\theta /2), \\
C &=&-3\gamma M_{{\rm s}}(\zeta _{y}\cos ^{2}\theta +\zeta _{z}\sin
^{2}\theta )/2.
\end{eqnarray*}
From (\ref{system}) and (\ref{ortho}) we obtain:

\begin{eqnarray}
&u_{11}=u_{21}=uU_{1}+vV_{1},\quad v_{11}=v_{21}=uV_{1}+vU_{1},&  \label{U-V}
\\
&u_{12}=-u_{22}=uU_{2}+vV_{2},\quad v_{12}=-v_{22}=uV_{2}+vU_{2},&  \nonumber
\end{eqnarray}
where

\begin{eqnarray}
U_{k} &=&\sqrt{\frac{\Omega -(-1)^{k}B+\omega _{k}}{4\omega _{k}}},\quad v=-%
\frac{D}{|D|}\sqrt{\frac{A-\Omega }{2\Omega }},  \nonumber \\
V_{k} &=&(-1)^{k}\sqrt{\frac{\Omega -(-1)^{k}B-\omega _{k}}{4\omega _{k}}}%
,\quad u=\sqrt{\frac{A+\Omega }{2\Omega }},  \nonumber \\
\omega _{k} &=&\sqrt{[\Omega -(-1)^{k}B]^{2}-C^{2}},\quad \Omega =\sqrt{%
A^{2}-|D|^{2}}.  \label{uUvU}
\end{eqnarray}
For equilibrium in the cross track direction the two modes correspond to
coherent and fanning about the $z$ direction (similar to the modes
illustrated in Fig.1a and 1b, respectively). From Eq.(\ref{U-V}) we can see
that the ``weight'' of the second mode is always equal to zero: $%
W_{2}\propto (u_{12}+v_{12}+u_{22}+v_{22})\cos \theta =0$. Thus, the
two-mode model is reduced with some modification to the case of the
single-mode model.

\bigskip

This work was partly supported by matching funds from the Center for
Magnetic Recording Research at the University of California - San Diego and
CMRR incorporated sponsor accounts.

Figure captions.

Fig. 1.

Standing spin-wave modes (qualitatively) with odd (a, c) and even (b, d)
numbers of oscillation humps.

Fig.2

Film with two-cell discretization.

\end{document}